\documentclass[aps,showpacs,amsmath,amssymb,twocolumn,prb,floatfix,superscriptaddress,a4]{revtex4}
\usepackage[dvips]{graphicx}
\usepackage{amsmath}
\newcommand{\vecto}[1]{\mathrm{\textbf{#1}}}  %% for vectors

\begin{document}

%\preprint{}

\title{Controlled Operations in a Strongly Correlated Two-Electron Quantum Ring}

%\thanks{}
\author{E. Waltersson}
\affiliation{Atomic Physics, Stockholm University, AlbaNova, S-106
91 Stockholm, Sweden}

\author{E. Lindroth}
\affiliation{Atomic Physics, Stockholm University, AlbaNova, S-106
91 Stockholm, Sweden}

\author{I. Pilskog}
\affiliation{Department of Physics and Technology, University
    of Bergen, N-5020 Bergen, Norway}

\author{J.P. Hansen}
\affiliation{Department of Physics and Technology, University
    of Bergen, N-5020 Bergen, Norway}

\date{\today}

\begin{abstract}
We have analyzed the electronic spectrum and wave function
characteristics of a strongly correlated two-electron quantum ring
with model parameters close to those observed in experiments. The
analysis is based on an exact diagonalization of the Hamiltonian in a
large B-spline basis. We propose a new qubit pair for storing quantum
information, where one component is stored in the total electron spin
and one multivalued ``quMbit" is represented by the total angular
momentum. In this scheme the controlled NOT (CNOT) quantum gate is
demonstrated with near 100\% fidelity for a realistic far infrared
electromagnetic pulse.
\end{abstract}


\maketitle 

\section{Introduction}
Quantum gates based on entangled states have in recent times been
proposed in several exotic physical systems, e.g.~in ion
traps~\cite{zoller,molmer} and in cold Rydberg
atoms~\cite{lukin-dipoleblockade}. For a solid state realization, it
has been suggested to represent qubits through the electron spin
confined in so called quantum dots~\cite{divencincia}. Controlled
operations in a network of quantum dots have subsequently been
demonstrated~\cite{cellular-automata}, and more recently also achieved
in dot molecules~\cite{petta}.  A major challenge in such systems is
decoherence through interactions with the environment such as
hyperfine interaction with the surrounding bath of nuclear spins or
coupling to bulk phonon modes~\cite{taylor, meunier}. In this respect,
operations based on laser driven transitions between the involved
states~\cite{saelen,optical-dot-molecule,all-optical-cnot-dot} may
have certain advantages as they can be performed much more rapidly
then those induced by microwaves\cite{microwave_spec,microwave_oversight}.

Quantum dot experiments have now shown that manipulations of single
spins as well as state to state electronic transitions are feasible
and the technology is continuously improving~\cite{koppens,
  single-state-manipulation-2}. Progress regarding so called quantum
rings has developed in parallel, theoretically~\cite{simonin, benito,
  cliemente} as well as experimentally~\cite{fuhrer, fuhrer2}. A
qualitative understanding of the electronic structure is already well
established~\cite{suzanne,Energy_structure_qr,Gudmundsson} and studies of correlated few--electron
rings have been performed, see e.g.~\cite{chakra,saiga,ring_FIR,twoel_ring_Realistic_pot,ir_magnetic_ring}.

\begin{figure}
\includegraphics[width=\columnwidth]{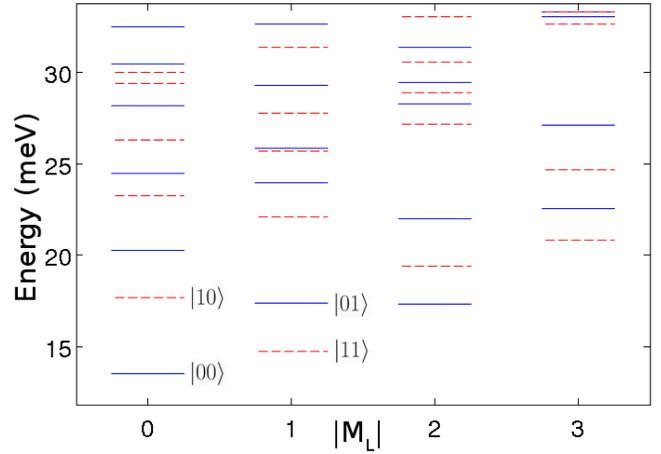}
 \caption{(Color online) Energy levels as functions of the absolute
   value of the total angular momentum $|M_L|$. The full blue (dark
   gray) lines correspond to singlets and the dashed red (light gray)
   lines to triplets. The states used to realize the CNOT gate are
   labeled using the notation $|S M_L \rangle$.\label{fig1}}
\end{figure}

Some progress towards quantitative operational quantum gates has
been made through suggestions for controlled persistent ring current
schemes~\cite{ring-gate-1}, and numerical demonstrations of fast
coherent control in a one-electron quantum ring~\cite{rasanen,
  luis}. In the present letter we have instead analyzed a strongly
correlated two-electron quantum ring. For the design of quantum
gates in the time domain, a characterization including both
electron-electron interactions and realistic system parameters is a
prerequisite, and the understanding of the spectrum of excited
states is of utmost importance. With such knowledge, it is possible
to design an electromagnetic pulse to optimize transitions, as
recently shown in the case of a two-electron quantum dot
molecule~\cite{saelen}. For this purpose, we perform an exact
diagonalization of the two-electron quantum ring Hamiltonian with
realistic model parameters. From the results, we characterize the
wave functions in terms of conserved quantum numbers, probability
densities and probability currents. Based on this analysis, we
propose a new form for quantum information storage and show that a
controlled two-bit operation can be performed with almost 100\%
fidelity. 

\section{Theory}

\subsection{One electron}
The Hamiltonian of one electron confined in a 2D quantum ring, modeled
by a displaced harmonic confinement rotated around the $z$-axis, can
be written
\begin{equation}
\label{singleparticlehamiltonian}
\hat{h}_s=
\frac{\hat{\vecto{p}}^2}{2m^*}
+\frac{1}{2}m^*\omega_0^2(r-r_0)^2,
\end{equation}
where $m^* = 0.067m_e$ is the effective mass of GaAs, $\omega_0$
corresponds to the potential strength, $r$ is the radial
coordinate and $r_0$ the ring radius. 

In polar coordinates, the radial equation then reads
\begin{equation}
\label{singleparticleradialeq}
\left [ \frac{\hbar^2}{2m^*}\left( -\frac{\partial^2}{\partial r^2} +
\frac{m_l^2}{r^2} \right)
+\frac{1}{2}m^*\omega_0^2(r-r_0)^2 - E \right]u_{nm_l}(r) = 0, 
\end{equation}
where $m_l$ is the angular quantum number and $u_{nm_l}(r)$ is the
radial function. 

Throughout this work $r_0 = 2$ a.u.$^* \approx 19.6$ nm and
$\hbar\omega = 10 $meV have been used. These potential parameters
correspond well to what has been measured in
experiments~\cite{Lorke_2}. Note that we here use the abbreviation
a.u.$^*$ for effective atomic units, i.e. atomic units that have
been rescaled with the material parameters $m^*$ and $\epsilon_r$. 

The one--particle wave functions are then found by the single particle
treatment described in section II A of
Ref.~\cite{waltersson_dot1}. Here, however, we use a knot sequence
that is centered around $r_0$ and where the knot points are
distributed using an $\arcsin$ function.

\subsection{Two electrons}
The two--electron Hamiltonian is written as
\begin{equation}
\hat{H}_0 = \hat{h}_s^1 + \hat{h}_s^2 +
\frac{e^2}{4\pi\epsilon_r\epsilon_0 \hat{r}_{12}},
\end{equation}
where $\epsilon_r = 12.4$ (GaAs). The corresponding eigenvalues are
found by exact diagonalization in a basis set consisting of
eigenstates to $\hat{h}_s^1 + \hat{h}_s^2$, as explained in
Ref.~\cite{waltersson_dot1}. The basis set is truncated at $n = 13$
and $|m_l|=12$, yielding matrix sizes in the order of $5000\times
5000$.

\smallskip
To visualize the many body states we calculate the probability
density $\rho$ and probability current $\vecto{j}$ and integrate out
the coordinates of one of the electrons,
\begin{eqnarray}
\rho(\vecto{r}_1) &=& \int d\vecto{r}_2  |\Psi(\vecto{r}_1,\vecto{r}_2)|^2
\label{prob_density}
\\
\vecto{j}(\vecto{r}_1) &=& \Re \left[
\int d\vecto{r}_2 \Psi^*(\vecto{r}_1,\vecto{r}_2)
\left(
-\frac{i \hbar}{m^*} \nabla_1 \Psi(\vecto{r}_1,\vecto{r}_2)
\right)
\right].
\label{prob_current}
\end{eqnarray}
Similarly we calculate the \emph{relative} probability density
$\tilde{\rho}$ and the \emph{relative} probability current $\tilde{j}$
by the coordinate transformation $\phi_1\rightarrow \phi_{rel}=\phi_1-\phi_2$
\begin{eqnarray}
\tilde{\rho}(r_1, \phi_{rel}) &\equiv& \rho (r_1, \phi_1-\phi_2)  
\label{rel_density}
\\
\tilde{\vecto{j}}(r_1, \phi_{rel}) &\equiv& \vecto{j}(r_1, \phi_1-\phi_2).
\label{rel_current}
\end{eqnarray}
Since there is no preferred angle $\phi$ this is equivalent to freezing
one electron at $\phi=0$ and calculating the probability density
(current) of the other one.

\begin{figure}
\includegraphics[width=8.6 cm]{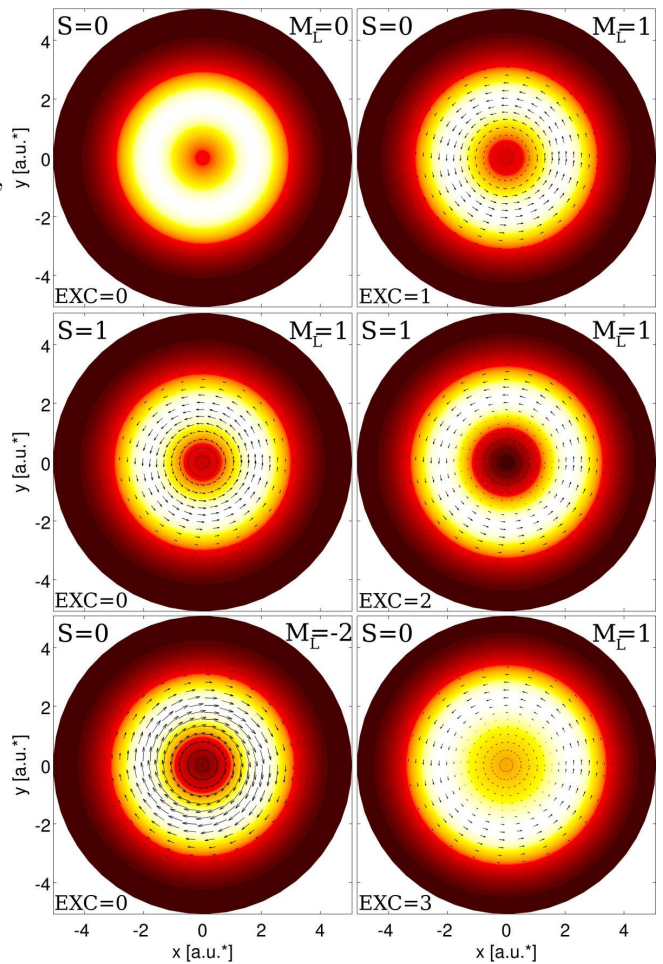}
 \caption{(Color online) Probability densities and currents of the two
   electron ring plotted in the coordinates of the first electron, see
   eqs.~(\ref{prob_density}) and (\ref{prob_current}). The left column
   depicts the lowest energy states (EXC$=0$) for $M_L = 0, 1$ and
   $-2$ counting downwards. The right column depicts the first, second
   and third excited state (EXC$=1, 2$ and $3$) with $M_L=1$, also
   counting downwards. Here the ring radius is set to $r_0 = 2$ a.u.$^*
   \approx 19.6$ nm.
\label{fig2}}
\end{figure}

\begin{figure}
\includegraphics[width=8.6cm]{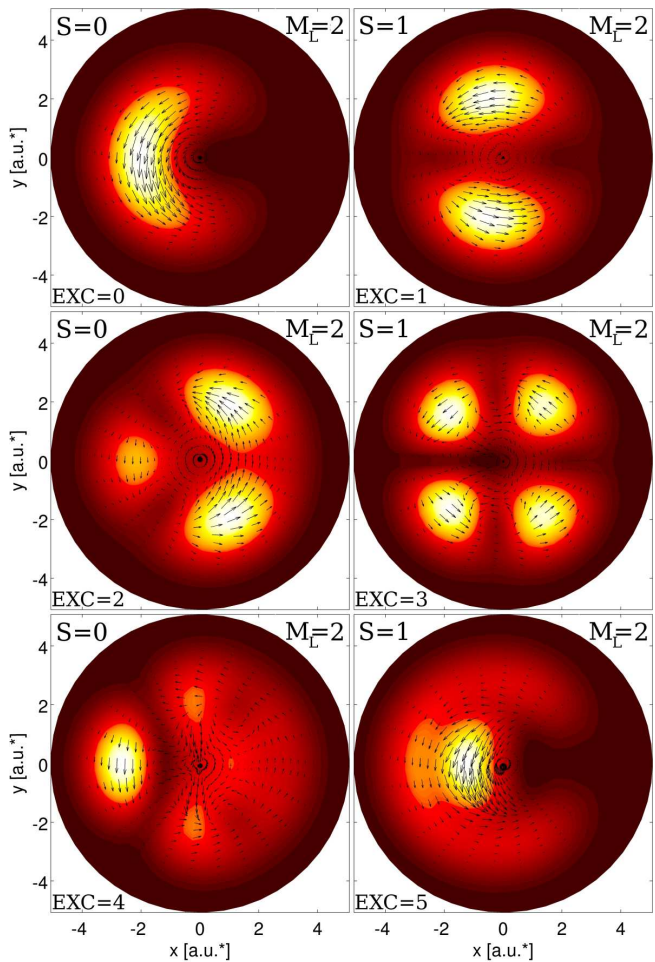}
 \caption{(Color online) Relative probability densities and currents,
   see eqs.~(\ref{rel_density}) and (\ref{rel_current}), of the $M_L =
   2$ system. The left column depicts singlets and the right triplets,
   starting with the lowest lying singlet (triplet) at the top left
   (right) corner, continuing with the first excited singlet (triplet)
   in the middle left (right) panel and so on, compare
   Fig\ref{fig1}. The label EXC$=0$ is used for ground states, EXC$=1$
   for the first excited state etc. Here the ring radius is set to
   $r_0 = 2$ a.u.$^* \approx 19.6$ nm.
\label{fig3}}
\end{figure}

\section{Structure of the strongly correlated quantum ring}
Fig.~\ref{fig1} shows the energy level scheme. The full blue (dark
gray) lines represent spin singlets ($S=0$) and the dashed red (light
gray) lines spin triplets ($S=1$). We observe large singlet--triplet
splittings, e.g.~between the first and second excited $M_L=0$ states,
caused by the electron-electron interaction representing $\sim 30\%$
of the energy.

Fig.~\ref{fig2} depicts the probability densities and probability
currents, eqs.~(\ref{prob_density}) and (\ref{prob_current}), for a
set of different states. The lowest energy state for each
$M_L$--symmetry are all ring formed and the expected properties;
increasing probability currents with increasing
$|M_L|$ and sign dependent direction of the probability current, are
clearly shown. Moreover, both the first and second excited $M_L = 1$
states are ring formed while the third excited state shows a more
dot-like behavior with a relatively large probability density at the
center of the system.

In Fig.~\ref{fig3} the relative probability densities and relative
probability currents, eq.~(\ref{rel_density}) and (\ref{rel_current}),
of the six lowest lying $M_L = 2$ states are shown. The lowest lying
state has one relative current density peak, the first excited state
has two peaks etc, up to the third excited state. These vibrational
excitations are expected in a quantum ring~\cite{suzanne}.  The fourth
and fifth excited state, however, do not continue this quantum ring
pattern, indicating that these more energetic states are more
dot-like. For the relative probability current, however, signs of
deviation from ring behavior are seen earlier. While for a large (or
quasi 1D) ring, the radial component of the relative probability
currents would approach (be) zero, the currents here show a rich
structure. Already at the first excited state, and even more clearly
in the higher lying states, we see complete departure from this
circular shape. Even probability current vortices can be seen,
i.e. between the peaks in the third excited state. Hence, we are here
in a region of strongly correlated electrons that still exhibit
ring-like behavior.

\section{The Controlled NOT operation}
The conservation of the total spin ($S$) and angular momentum ($M_L$) suggests
the possibility to apply these two variables for storage of quantum
information, such that one qubit is represented by $S$ and another
multivalued ``quMbit" is stored in $M_L$. Single qubit operations
involving the latter may then be performed by carefully optimized spin
conserving electromagnetic interactions \cite{saelen}. Controlled spin
manipulation will in general be more complicated but can be performed
by various schemes involving inhomogeneous magnetic
fields~\cite{fuhrer, fuhrer2}. The control of the time-scale of the
two single-qubit operations will be a matter of magnetic field
inhomogeneity vs. size of the quantum ring: Experimentally, the degree
of inhomogeneity is allowed to increase with the ring-size, thereby
decreasing the spin flip period.  The transition period between
electronic states, that for small radii are much shorter, will on the
other hand increase with the ring radius. At some point, both of these
single qubit operations can be performed at comparable time scales.

\begin{figure}
\includegraphics[width=8.6cm]{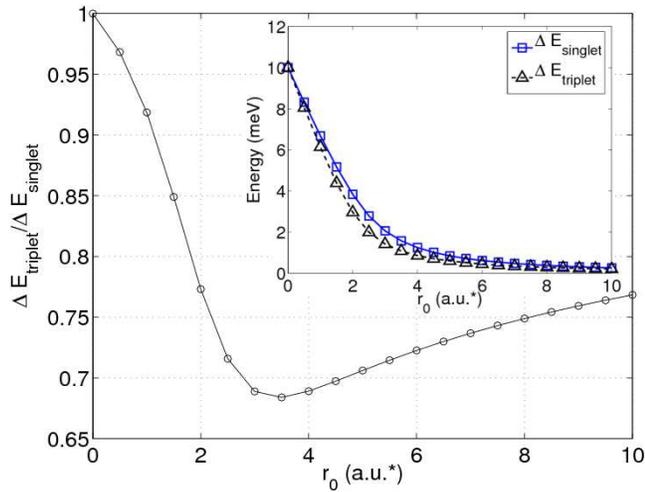}
 \caption{(Color online) The quotient between $\Delta
   E_{\textrm{triplet}}= E_{|10\rangle} - E_{|11\rangle}$ and $\Delta
   E_{\textrm{singlet}} = E_{|01\rangle} - E_{|00\rangle}$ (see
   Fig.\ref{fig1}) as a function of the ring radius $r_0$. The insert
   shows the absolute value of $\Delta E_{\textrm{triplet}}$ and
   $\Delta E_{\textrm{singlet}}$ as a function of the same. Here the
   effective Bohr radius $=1$ a.u* $\approx 9.8$ nm.
\label{fig4}}
\end{figure}

The critical and remaining question is thus whether two-qubit
operations can be performed in such a way that a change of an initial
$|S, M_L \rangle$ can take place for a conditional value of $S$. The
relatively strong electron-electron interaction can here play a
constructive role as it changes the electronic energy shift between
internal singlet and triplet states, cf. Fig.~\ref{fig1}. In this way,
the ring size can be used as a parameter to tune the energy
spectrum. Fig.~\ref{fig4} depicts the quotient between $E_{|10\rangle}
- E_{|11\rangle}$ and $E_{|01\rangle} - E_{|00\rangle}$ (see
Fig.\ref{fig1}) as a function of the ring radius $r_0$. Starting at
unity for $r_0=0$, this quotient decreases to a minimum at $r_0
\approx 3$ a.u.* and then increases again for larger ring radii. To
realize a conditional operation we want this quotient to be as far
from unity as possible. However, to protect the CNOT against
decoherence we want the absolute value of both energy differences to
be large. These energy differences decrease monotonically with the
ring radius, see insert of Fig.~\ref{fig4}, yielding better protection
for smaller radii. Weighting these two things together, a ring
radius $\sim 2$ a.u.* seems a close to optimal choice.

\begin{figure}
\includegraphics[width=8.6cm]{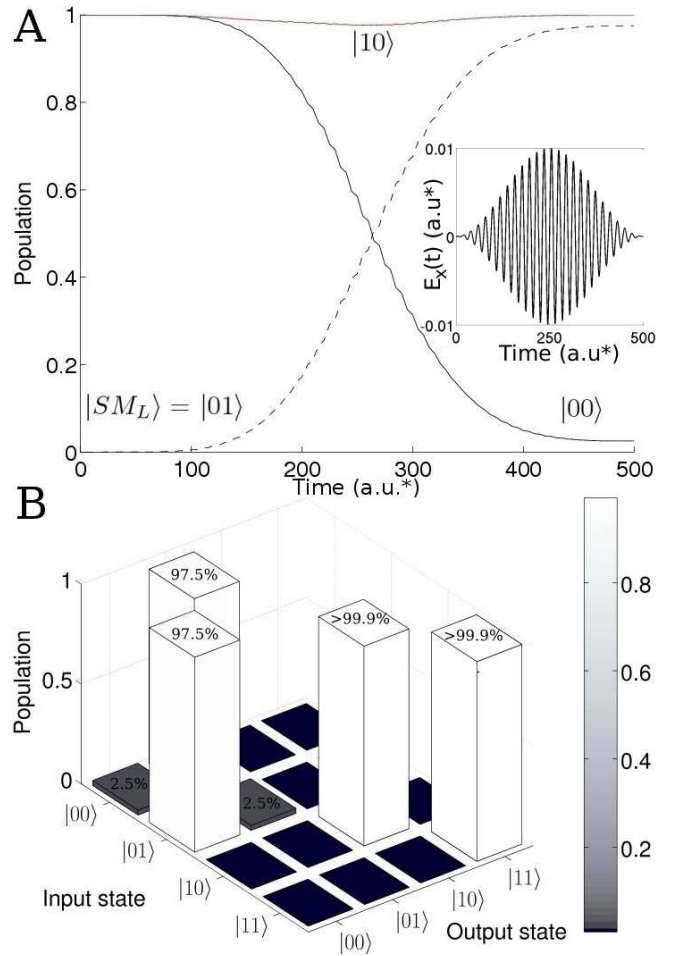}
 \caption{\label{fig5}(Color online) Upper panel: The time development
   of the populations of the different states when the central
   frequency, $\omega_L$, of the pulse corresponds to the energy shift
   between the the two lowest states in the singlet system, $\Delta
   \varepsilon_{|S M_L\rangle} = \varepsilon_{|0 1\rangle}-
   \varepsilon_{|0 0\rangle} \approx 3.8$ meV implying a laser
   frequency of $0.9$ Thz. The population of the state $| S M_L
   \rangle = | 0 0 \rangle$ is almost completely transferred to $ |0 1
   \rangle$, while the population of the $ | 1 0 \rangle$ is seen to
   be nearly constant. The insert shows the $x$-component of the
   electromagnetic pulse. Lower panel: The truth table shows just a
   small amount of unwanted population. The pulse length was chosen to
   be $T=500$ a.u* $\approx 28$ps. }
\end{figure}

\subsection{The interaction with the electromagnetic field}
We now examine transitions induced by a circularly polarized
electromagnetic pulse $ {\bf E}(t) = E(t) [ \cos{(\omega_L t)}
\hat{\vecto{x}} \pm \sin{(\omega_L t)} \hat{\vecto{y}}]$ where $\omega_L$
is the central frequency. The electric dipole interaction then couples
neighboring $M_L$ states ($\Delta M_L = 1$). The envelope $E(t)$ is taken as, $E(t) = E_0
\sin^2 \left( \pi t/T \right)$, which defines a pulse that lasts from
$t=0$ to $t=T$. Here we set $T = 500$ a.u.$^* \approx 28$ ps and
$E_0\approx 0.01$ a.u.$^*$ corresponding to an intensity $\sim
2.4\cdot 10^2$W/cm$^2$.  We solve the time dependent Schr\"odinger
equation in the full basis of eigenstates obtained from the previous
diagonalization. The Hamiltonian then becomes
\begin{equation}
\hat{H}(t) = \hat{H}_0 + e \vecto{E}(t)\cdot (\vecto{r}_1 + \vecto{r}_2).
\end{equation}
It is then readily shown that the time dependent Schr\"odinger
equation can be written as a coefficient equation including the
transition matrix elements from (the two particle) state $|j\rangle$
to state $\langle i|$
\begin{equation}
\dot{c}_{i}(t) = -i\sum_{j} c_{j}(t)  e\vecto{E}(t) \cdot \langle i |(
\vecto{r}_1 + \vecto{r}_2) | j \rangle.
\end{equation}

\subsection{The realization of the CNOT}
 Fig.~\ref{fig5}A depicts the time development of the populations of
 the different states when the pulse central frequency, $\omega_L$,
 corresponds to the energy shift between the two lowest states in the
 singlet system $\Delta \varepsilon_{|S M_L\rangle} = \varepsilon_{|0
   1\rangle }- \varepsilon_{|0 0 \rangle} \approx 3.8$ meV. The
 driving laser frequency would then be $\omega_L/2\pi \approx 0.9$
 THz. With the initial state being $| S M_L \rangle = | 0 0 \rangle$,
 we observe a nearly complete transition to $| 0 1 \rangle$, with a
 small amount of unwanted population. Also shown is the time
 development of the population of an initial $ | 1 0 \rangle$ which is
 seen to be nearly constant.  Thus a CNOT is realized, as clarified in
 the truth table of Fig.~\ref{fig5}B. It clearly depicts how the
 electromagnetic pulse transfer $\sim 97.5\%$ of the spin singlet
 population while leaving $> 99.9\%$ of the triplet population
 unchanged.

\section{Decoherence, advantages and possible improvements}
The natural life time of the excited state is dominated by phonon
relaxation and is at least a few orders of magnitudes longer than the
time for the induced transition studied here~\cite{Phonon_scatt}. As
shown recently, the pulse induced transition time may then be
decreased by a factor 10-100 through quantum optimization field
control methods optimized to almost any fidelity~\cite{saelen, luis}.
Transition times will also be drastically reduced by increased
confinement strength which will lead to considerably higher central
frequencies, $\omega_L$. In the present work, however, we wanted to
examine a region of confinement strengths that already is accessible in
experiments~\cite{Lorke_2}.

The present proposal has certain advantages compared to single quantum
dot qubit systems\cite{koppens}. First, the energy difference between
the spin states prohibit unwanted transitions up to the order of 10
Kelvin. In the quantum molecule qubit, the energy differences are
typically $10^3$ times smaller\cite{taylor}. Furthermore, by taking
advantage of a long array of accessible $M_L$ levels, one may develop
much more powerful algorithms for certain operations than systems with
only two levels. 

\section{Summary and conclusions}
In conclusion we have shown that quantum controlled operations
defining a conditional two-bit transition can be realized in a
two-electron quantum ring. This has been achieved through a detailed
analysis of energy levels and properties of the wave functions. The
electron-electron interaction has been utilized to effectively store
quantum information simultaneously in the total spin and total angular
momentum. We have shown that with realistic model parameters it is
possible to find a regime where the singlet and triplet splittings
differ such that an electromagnetic pulse can transfer
population of one spin state to a higher energy level, while leaving
the population of the other spin state intact. This opens for
a new type of solid state quantum information devices, for which
quantum rings are shown to be promising candidates.

\end{document}